\documentclass[aps,prb,twocolumn,groupedaddress,showpacs]{revtex4-1}
\usepackage{color}
\usepackage{amsmath,amsthm,amssymb}
\usepackage{graphics}
\usepackage{graphicx}
\usepackage{hyperref}
\usepackage{subfigure}
\usepackage{float}

\begin{document}

\title{Exchange constants and spin waves of the orbital ordered,
non--collinear spinel MnV$_{2}$O$_{4}$}
\author{R.~Nanguneri, S. Y.~\ Savrasov}
\affiliation{Physics Department, University of California, Davis, California 95616, USA}

\begin{abstract}
We study the exchange constants of MnV$_{2}$O$_{4}$ using magnetic force
theorem and local spin density approximation\ of density functional theory
supplemented with a correction due to on--site Hubbard interaction $U$. We
obtain the exchanges for three different orbital orderings of the Vanadium
atoms of the spinel, two sizes of trigonal distortion, and several values
of Coulomb parameter $U$. We then map the exchange constants to a Heisenberg
model with single--ion anisotropy and solve for the spin--wave excitations
in the non--collinear, low temperature phase of the spinel. The single--ion
anisotropy parameters are obtained from an atomic multiplet
exact--diagonalization program, taking into effect the crystal--field
splitting and the spin--orbit coupling. We find good agreement between the
spin waves of one of our orbital ordered setups with previously reported
experimental spin waves as determined by neutron scattering. We can
therefore determine the correct orbital order from various proposals that
exist in the literature.
\end{abstract}

\maketitle

\section{Introduction}

Transition metal oxides (TMO) are a class of solid--state materials that
exhibit a rich variety of physical phenomena\cite{tokura00}. Among them,
magnetic cubic spinels AV$_{2}$O$_{4}$ have recently attracted much
attention due to geometrically frustrated corner sharing tetrahedral network
formed by the V atoms (also known as a pyrochlore lattice)\cite{plum87}. An
interesting example is represented by MnV$_{2}$O$_{4}$ which is the spinel
having additional magnetic Mn ions. It exhibits an orbital ordering (OO)\
that occurs at finite $T$ as a thermal phase transition: At room
temperature, crystalline MnV$_{2}$O$_{4}$ is a cubic paramagnet (PM)
where Mn sites occupy the centers of oxygen tetrahedra (MnO$_{4}$ units),
while V sites occupy the centers of oxygen octahedra (VO$_{6}$ units) which exhibit slight trigonal
distortions consistent with the $Fd\overline{3}m$ cubic symmetry. As $T$
is lowered there occur two phase transitions: [1] A magnetic transition at $T_{F}=56$ K from the
high--$T$ PM phase to a cubic ferrimagnetic (FEM) phase, with the Mn and V
moments anti--aligned; [2] followed by a second transition at $T_{S}=53$ K to a tetragonal, non--collinear FEM with orbital ordering of $V^{3+}$ $3d^{2}$ electrons\cite{gar08}.
The orbital ordered phase is accompanied by a reduction of the V magnetic moments due to the
formation of the electron orbital moment (finite orbital angular momentum).
The orbital moment, $m_{o}\approx 0.34$, is anti--aligned with the spin
moment, $m_{s}\approx 1.65$, giving the total moment of $m\approx 1.31$ \cite%
{gar08}. The reduced value of V moment has been reproduced by an earlier
first--principles work in Ref. \onlinecite{sarkar09}, and is explained by
the spin--orbit coupling (SOC)\ on the V $3d^{2}$ which generally favors
anti--alignment of spin and orbital angular momenta for $T$ below the energy
scale of SOC \cite{plum87}.

The local tetrahedral and octahedral coordination of the Mn and V sites results in the crystal--field (CF) splitting of their 5-fold $3d$ orbital degeneracy. Tetrahedrally coordinated Mn has an $e_{g}$ lower in energy than $t_{2g}$, while the splitting is opposite for octahedrally coordinated V. Inter--electron Coulomb interactions and exchange anti--symmetry lead to Hund's rule splitting of
up and down spins, which is greater than the CF splitting. In the
stoichiometric crystalline environment, Mn has an outer shell high--spin $%
S=5/2$ configuration of $3d^{5}$ and a valence of $+2$: all 5 up--spin $3d$
orbitals are occupied giving $L_{z}=0$ (quenched total orbital moment), and the down spin ones are empty.
V has a valence of $+3$, an outer shell configuration of $3d^{2}$, and $S=1$: in this
case, 2 electrons must occupy the 3 $t_{2g}$ orbitals. In
the high temperature cubic phase, these latter three are nearly degenerate, while
in the low temperature tetragonal phase, where the unit cell is slightly
compressed along the $c$--axis, the $xy$ is lowered in energy while $yz$, $%
zx$ remain degenerate. Thus, in the tetragonal (low--$T$) phase, one
electron on V occupies the $xy$, and the second electron has the freedom to
occupy either $yz$, or $zx$. Unlike Mn, the orbital angular momentum of V is not fully quenched: The partial
occupation of the $yz$ and $zx$ gives an effective orbital angular momentum $%
L=1$ for V. The fact that $L\neq 0$ implies that there maybe non--negligible
effects of SOC in the V atoms \cite{plum87}. Further, this is a hint that the $yz$ and $zx$
could form complex linear combinations of one--electron states if it happens that $L_z=\pm1$, since only such a
complex state can have a non--zero $L_z$. The freedom of the
second electron of V to occupy $yz$, $zx$, or some linear combination of the
two gives rise to the possibility of long--range orbital order in the low--$T$ phase.

\begin{figure}[tbph]
\includegraphics[width=0.5\columnwidth]{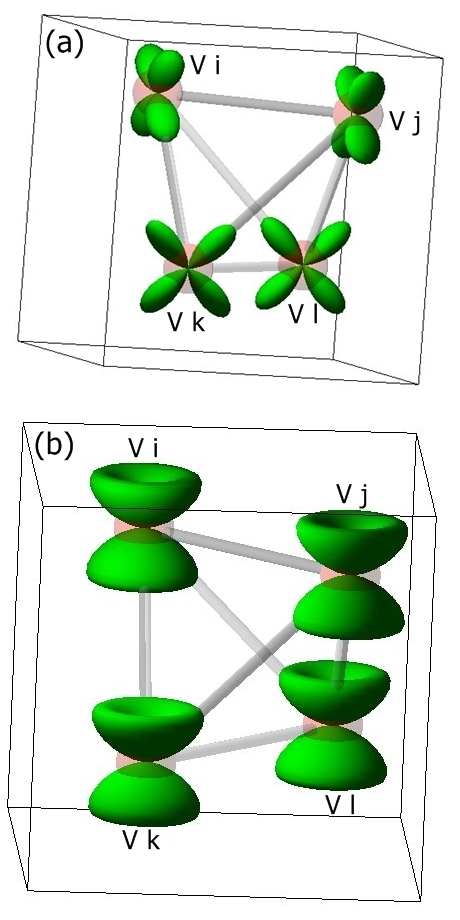}
\caption{(a) Schematic illustration of the \textit{initial} real antiferro--orbital order
of the type I (AFOO I) with $L=0$ on the four corners of the V tetrahedron.
The lower and upper horizontal bonds are in the $ab$ plane. The red spheres
are the V atoms. The lower $ab$ plane has $yz$ orbitals occupied on all V,
while the upper $ab$ plane has $zx$ occupied on all V.
(b) Schematic illustration of \textit{initial} ferro orbital order on all four
corners of the V tetrahedron where an electron occupies the same real linear
combination of $yz$ and $zx$ on all V sites. Note that the self--consistent
solution breaks this symmetry and results in an electron occupying
alternately $\protect\psi _{+}=(\protect\psi _{yz}+\protect\psi _{zx})/%
\protect\sqrt{2}$ and $\protect\psi _{-}=(\protect\psi _{yz}-\protect\psi %
_{zx})/\protect\sqrt{2}$ along the $c$--axis. We refer to this order as
antiferro--orbital order of type II\ (AFOO II).
The indices {\rm i}, {\rm j}, {\rm k}, {\rm l} denote the inequivalent V sites in the FCC primitive cell. }
\label{fig1}
\end{figure}

Two simple choices has been proposed for the orbital ordering in this spinel and both have been studied theoretically in mean field models. One is
the Antiferro--Orbital Order (AFOO) with alternate occupation of the $yz$
and $zx$ along the c--axis, i.e.: the same orbital is occupied in a given $%
ab$--plane but the other orbital is occupied in the adjacent planes above
and below\cite{tsun03,gar08,suzuki07}, as shown in Fig.~\ref{fig1}(a). This order has the space--group
symmetry $I4_{1}/a$. The second is the Ferro--Orbital Order (FOO) where the
same orbital is occupied on all V--atoms\cite{adachi05}, giving the space--group $I4_{1}/amd$, as shown in Fig.~\ref{fig1}(b). In the latter, if the orbital order is a complex linear combination of $yz$
and $zx$ there will be a non--zero orbital angular momentum
and a magnetic moment associated with it\cite{tchern04}. Spin--orbit
coupling can stabilize the finite orbital moment, since the energy is lower
for anti--parallel alignment of $\vec{L}$ and $\vec{S}$.

In both of the above proposals, trigonal distortion of the VO$_{6}$
octahedra in the low--$T$ phase is not taken into account, but it is known to be large in MnV$_{2}$%
O$_{4}$ as compared to other vanadates. While a slight trigonal distortion is present even in the high--$T$ cubic phase, there is a qualitative symmetry--lowering change and an increase in this distortion in the low--$T$ phase which lifts the residual
degeneracy between $yz$, $zx$, and of the $e_{g}$ manifold, and combined with the tetragonal distortion results in the mixing all 5 $3d$ orbtials.
In this case, the above OO proposals are not necessarily correct as these
assume degeneracy between the $yz$, $zx$ orbitals. This low--$T$ trigonal distortion
has indeed been observed in the previous first--principles work\cite{sarkar09}
that used local spin density approximation (LSDA) of density functional
theory (DFT)\ \cite{DFTBook}\ supplemented by the correction due to on--site
Hubbard interaction $U$ \cite{AnisimovLDA+U}\ for correlation strengths $U>2$
eV. In that work, in addition to a tetragonal relaxation (compression) along
the $c$--axis, structural relaxation of the O positions is performed and a
trigonal distortion of the VO$_{6}$ octahedron with a concomitant lowering
of symmetry from $I4_{1}/amd$ to $I4_{1}/a$ is found.
By projecting the converged density onto an atomic orbital basis using so
called N--th order muffin--tin orbital (NMTO)\ downfolding \cite%
{AndersenNMTO}, the authors of Ref.~\onlinecite{sarkar09} find a different
electron occupation order from the ones proposed above, namely, the first
electron occupies the lowest energy eigenstate, and the second occupies the
next higher energy eigenstate. The $3d$ energy eigenstates are the same on all V sites, but rotated
alternatively by $45^{\circ }$ along the $ab$--chains due to the staggered
trigonal distortion. Thus, the same orbitals are occupied on all V sites,
akin to the FOO, but nevertheless the space-group symmetry is $%
I4_{1}/a$ expected of AFOO due to the trigonal distortion.

The low--$T$ magnetic excitations of the compound have been mapped along
high--symmetry directions using inelastic neutron scattering\cite%
{chung08,gar08}. At the $\Gamma $ point, these excitations are gapped for
the acoustic modes, indicating the presence of single--ion anisotropy, which
essentially occurs due to the interplay between SOC and crystal--fields\cite%
{alders01}. In Ref. \onlinecite{chung08}, the authors start with a
nearest--neighbor Heisenberg Hamiltonian including the anisotropy term and
calculate spin--wave spectra and corresponding eigenmodes using linear
spin--wave theory (LSWT) for the non--collinear, tetragonal phase. By
fitting the spectrum to inelastic neutron scattering data, they were able to determine the exchange
couplings between Mn--V, V--V in $ab$--plane, and V--V between $ab$--planes
along the $c$--axis. They find all exchanges to be AFM with the following
values:

\begin{itemize}
\item $J_{\rm Mn-V} = -2.82$ meV

\item $J_{\rm V-V}^{ab} = -9.89$ meV

\item $J_{\rm V-V}^{c} = -3.08$ meV
\end{itemize}

The authors point out the interplanar coupling between V atoms, $J_{\rm V-V}^{c}$, along the $c$--axis
is unusually large for AFOO because such an alternate orbital occupation in
the vertical direction would yield negligible orbital overlap, and would
also be ferromagnetic (wrong sign) by the Goodenough--Kanamori rules\cite%
{good63}. The alternate proposal, FOO, would be consistent with these
results, but would have the wrong symmetry, $I4_{1}/amd$. The symmetry group
of this spinel vanadate has been established conclusively as $I4_{1}/a$ by a
synchrotron x--ray study\cite{suzuki07} which supports AFOO, but contradicts
with the large value of $J_{\rm V-V}^{c}$.

A possible resolution of this puzzle is that trigonal distortion has been ignored in these simple proposals. With
trigonal distortion, we expect a more complex orbital ordering which has the
requisite symmetry $I4_{1}/a$ and would give the observed (or fitted) $%
J_{\rm V-V}^{c}$ along the $c$--axis\cite{chung08}. This is exactly what has
been found in the ab--initio work of Ref.~\onlinecite{sarkar09}. Their
physical picture has received some support by a recent $^{51}$V NMR
work of Ref.~\onlinecite{baek09} and by analytical model of Ref. %
\onlinecite{Perkins}.

In this work, we report our study of MnV$_{2}$O$_{4}$ based on the LSDA+$U$
method and using linear muffin--tin orbital (LMTO) basis set to solve the
electronic structure problem \cite{Andersen1975,Savrasov1996}. We calculate
the pair--wise interatomic magnetic exchange interactions ($J$) between all
magnetic atoms using linear response theory and magnetic force theorem \cite%
{liech87,wan06}, including the single--ion anisotropies ($D$) for Mn and V
found by the exact diagonalization procedure\cite{alders01}. We then use the
obtained $J$ and $D$ as parameters in a Heisenberg Hamiltonian with
anisotropy to derive the spin--wave spectra in a semiclassical
approximation. We explore three initial orbital ordering scenarios: [1]
Antiferro, [2] Ferro, and [3] Complex ferro + SOC in the density matrix of
the $3d$ shell of V to see how they affect the obtained exchange
interactions. We also performed non--collinear magnetic electronic structure
calculations.

In our low-$T$ tetragonal structures, we explore the effects of two types of trigonal distortions of the ${\rm VO}_6$ octahedra: A small trigonal distortion, of order $2\%$ of the undistorted structure, with $I4_1/amd$ symmetry; and a larger trigonal distortion, of the type used in the relaxed structure of Ref.~\onlinecite{sarkar09} (about $10\%$ of the undistorted structure), with an $I4_1/a$ symmetry.
We find that the $J^{\prime}$s depend on both the size of trigonal distortion and Coulomb parameter $U$; we are thus faced with a two-parameter `trigonal--distortion/Coulomb--$U$' space within which to search for a good match between experimental and theoretical $J^{\prime}$s. We find that SOC complex ferro--orbital order give $J^{\prime}$s which best match the experimental ones for small trigonal distortion and low-$U$, and also for larger trigonal distortion and higher-$U$.

Our paper is organized as follows. We begin with a discussion of the
proposed orbital orders and their electronic structures in Section II. We
present our results for exchange interactions and comparisons with
experiment in Section\ III. We end with the conclusions in Section IV.

\section{Proposed Orbital Orders and their Electronic Structures}

We have done LSDA+$U$ calculations to model the electronic structure for all
three thermodynamic phases of MnV$_{2}$O$_{4}$. We describe our results in
the following subsections for the $T=0$ phase only since this is the phase
which exhibits orbital ordering and non--collinear magnetism. Our results
for the other finite--$T$ phases may be found in Ref.~\onlinecite{thesis}.
For the magnetic phases we use the same values of $U$ and $J_H$ for both Mn
and V correlated $3d$ shells. The use of the same $U$ on Mn and V is
justified because these elements have atomic numbers 25 and 23, and are thus
expected to have similar interaction strengths\cite{sarkar09}. The Coulomb
and exchange parameters in the solid state are generally screened, and hence
reduced by a considerable amount from their bare atomic values\cite{miyake08}%
. The structural parameters for all three phases are taken from experiment%
\cite{gar08}: In the cubic phase, the lattice constant is $16.0746$ a.u.,
and in the tetragonal phase it is $16.12$ a.u. with a small tetragonal
distortion ratio of $\frac{c}{a}=0.98$.

The non--collinear orbital ordered phase occurs when the temperature is
reduced below $T_{S}=53$ K. This phase transition results simultaneously in:
[1] a structural transition from cubic to tetragonal; [2] the canting of V
moments from a \textit{collinear} ferrimagnetic (FEM) to a $\mathbf{q}=0$
\textit{non-collinear} FEM spin order with non--zero components in the $ab$%
--plane; and [3] a long--range orbital order in the V $t_{2g}$ manifold. We
model the electronic structure of this phase using LSDA+$U$ method with $U=5$
eV and $J_H=1$ eV, but starting the self--consistency loop after imposition of
the initial orbital order(s) in the Hubbard--$U$ density matrix (further
described below), along with tetragonal distortion and two different
magnetic configurations: [1] \textit{collinear}, as in the intermediate
phase, and [2] \textit{non--collinear}, which is in fact the correct
magnetic order for this phase. The converged charge density for the low--$T$
\textit{collinear} calculation was used as the initial charge density for
the correct low--$T$ \textit{non-collinear} calculation. The orbital order
that is finally obtained after reaching the self--consistency is taken to be
the correct metastable solution within this approximation and specified
initial condition(s).

We initialize the V $3d$ density matrix to a particular orbital order by
specifying orbital occupation numbers in the atomic basis. This means we
initially specify only the diagonal components (occupation numbers) $\langle
n_{xy\uparrow }\rangle $, $\langle n_{yz\uparrow }\rangle $, $\langle
n_{zx\uparrow }\rangle $ of the density matrix for all four V atoms' $3d$
shells and set the off--diagonal elements to zero. The full complex density
matrix in the atomic basis is $\langle n_{m\sigma ,m^{\prime }\sigma
^{\prime }}\rangle $ (where $m,m^{\prime }$ and $\sigma ,\sigma ^{\prime }$
are the $3d$ orbital and spin indices respectively) and includes
off--diagonal components as well. As a result of the electron--electron
interactions, during the self--consistent cycle non--zero off--diagonal
components of the density matrix develop (since the interactions mix the
single--particle $3d$ orbitals at the Hartree--Fock (HF) mean--field level).
This means the true occupied orbitals are some linear combination of the
atomic basis functions. After convergence is reached, the final density
matrix, which is no longer diagonal in the $(m\sigma ,m^{\prime }\sigma
^{\prime })$ basis is diagonalized. The resulting eigenvectors and
eigenvalues give the `correct' single--particle HF wave functions and their
occupation numbers respectively. In the basis of these eigenfunctions, the
density matrix is once again diagonal, and its non--zero entries signify the
true orbitals which are occupied within the mean--field approximation of
LSDA+$U$. We are thus able to identify the orbital ordering that results after
convergence is attained. We describe the final orbital orders below for the
\textit{collinear} magnetic solutions only, since we find that the
electronic structures, and therefore the orbital orders, of the \textit{%
non-collinear} configurations are not significantly different from the
corresponding \textit{collinear} ones as discussed below.

\subsection{Anti-Ferro Orbital Order I: $I4_{1}/a$ symmetry}

\begin{figure}[tbp]
\includegraphics[width=1.0\columnwidth]{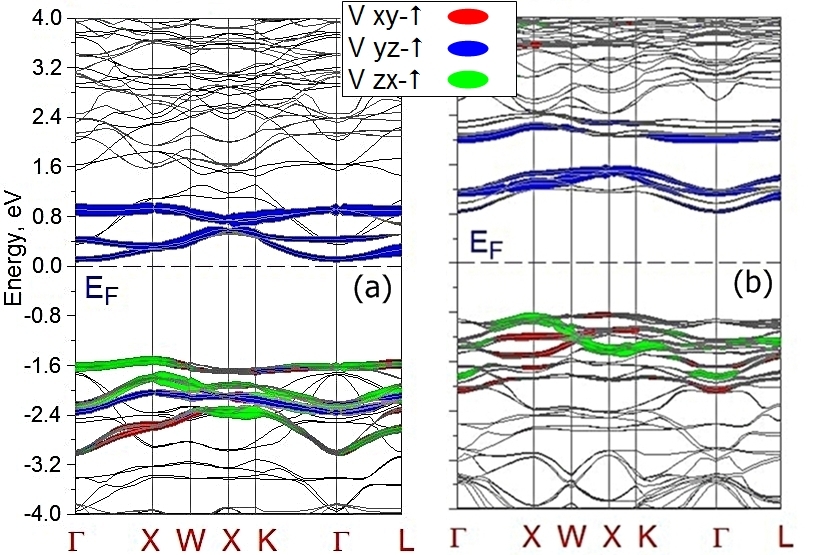}
\caption{(a) The V $t_{2g}$-$\uparrow$ bands for $U=5$ eV in the low--$T$
tetragonal phase with a \textit{collinear} ferrimagnetic spin configuration
and real antiferro--orbital order of the type I (AFOO-I) as discussed in
text.
(b) Bands for the same setup as in (a) but with a \textit{non-collinear} ferrimagnetic spin configuration.
In both panels, the partial characters of the $xy$-$\uparrow$, $yz$-$\uparrow$, $zx$-$\uparrow$ orbitals are for the sublattice {\rm i, j} V atoms. We see that due to the orbital ordering, the $zx$-$\uparrow$ is occupied while the $yz$-$\uparrow$ is somewhat less occupied. The occupations of these two orbitals are reversed for the sublattice {\rm k, l} V atoms on the adjacent parallel $ab$ planes along the $c$-axis. The sublattice indices are defined in Figs.~\ref{fig1},~\ref{fig5}. There is a band--gap of $E_{%
\mathrm{gap}}=1.67$ eV. }
\label{fig2}
\end{figure}

In the low--$T$ orbitally ordered phase, the tetragonal distortion occurs to
break the degeneracy of the $t_{2g}$ in both V and Mn. We first describe the case of small trigonal distortion. There is no orbital
freedom to place the electrons in the Mn $3d$. In the V, the energy of $xy$
gets lowered, so the first electron occupies $xy$. The second electron then
has the freedom to occupy the remaining degenerate orbitals $yz$ or $zx$.
Figure~\ref{fig1}(a) shows the initial orbital occupations with $I4_{1}/a$
symmetry. In this scenario, the second electron of V occupies either $yz$
or $zx$ alternately along the $c$--axis (antiferro OO), and the same orbital
within each $ab$ plane\cite{suzuki07,gar08}. (Each V chain within an $ab$
plane has the same orbital occupied.) The final converged density matrices of the V $3d$ subspace show that the converged orbital order is not the same as the initial order, but one which is similar to that found in Ref.~\onlinecite{sarkar09}. That is, when we rotate the $3d$ density matrix from the global tetragonal coordinate system to the local trigonal one, a rotation by $45^{\circ}$, we find the same set of eigenstates for all V atoms, and the lowest two of these states in energy are occupied. We label this order `AFOO-I', since it preserves the $I4_{1}/a$
symmetry.

The \textit{collinear} spin fat bands of V $t_{2g}$ electrons are shown in
Fig.~\ref{fig2}(a), and the same for \textit{non--collinear} spins in Fig.~\ref%
{fig2}(b). The occupations of the $t_{2g}$-$\uparrow$ bands, as shown by the partial characters,
reflects the converged orbital order, as well as the FEM spin configuration.
We also find that imposition of orbital order opens a gap of about $E_{%
\mathrm{gap}}=1.67$ eV at the Fermi level leading to an insulator state. The
qualitative features of the band structure and partial characters does not
change upon canting the V moments to the non--collinear configuration: The
band gap remains robust and the phase is still insulating.

Next we describe the case of large trigonal distortion with the space-group symmetry of $I4_1/a$. In this case we use $U=4.5$ eV, $J_H=1$ eV, along with muffin-tin sphere radii specified in Ref.~\onlinecite{sarkar09}. We start with an initial uniform orbital order in which the \textit{three} $t_{2g}-\uparrow$ are equally occupied, but the two $e_g$ are almost empty. We also start with a second initial orbital order consisting of equal occupations of all \textit{five} $3d-\uparrow$ orbitals. In both cases, we found that the converged density matrices, partial DOS, and fat-bands are identical with those of the calculation with small distortion. Our charge density on the V sites are alternately rotated within and between the V chains in the $ab$ plane, and as well, when we transform to the \textit{local} trigonal coordinate system at each of the V sites, we obtain the same single-particle wavefunctions, showing that the same orbitals are occupied on each V site, but are rotated alternately by 45$^{\circ}$ due to the trigonal distortion. Thus, both small and large trigonal distortions result in the \textit{same} orbital order, `AFOO-I'.

\subsection{Anti-Ferro Orbital Order II: $I4_{1}/a$ symmetry}

\begin{figure}[tbp]
\includegraphics[width=1.0\columnwidth]{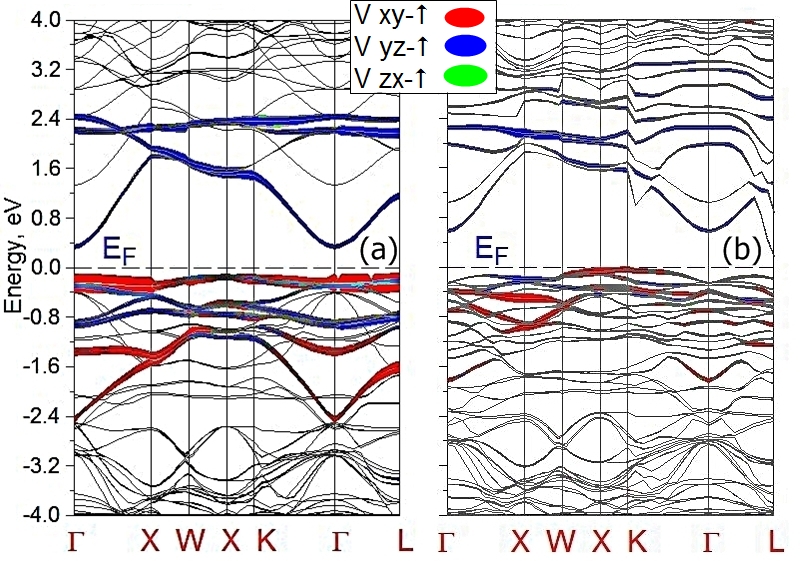}
\caption{(a) The V $t_{2g}$-$\uparrow$ bands for $U=5$ eV in the low--$T$ tetragonal
phase with a \textit{collinear} ferrimagnetic spin configuration and real
antiferro--orbital order of the type II (AFOO-II)\ as discussed in text.
(b) Bands for the same setup as in (a) but with a \textit{non-collinear} ferrimagnetic spin configuration.
In both panels, the partial characters of the $xy$-$\uparrow$, $yz$-$\uparrow$, $zx$-$\uparrow$ orbitals are for the sublattice {\rm i, j} V atoms. Note that the $yz$ and $zx$ partial characters have nearly identical dispersions due to their equal weight in the occupied orbital. There is again a band--gap for this orbital--order too.
}
\label{fig3}
\end{figure}

The next simplest \textit{initial} order has the second $t_{2g}$ electron
occupying the same real linear combination of $yz$ and $zx$ on all V sites,
with equal weight for both orbitals, see Fig.~\ref{fig1}(b). This initial order
has $I4_{1}/amd$ symmetry, and we implement only the small trigonal distortion. The real linear combination implies that the
orbital angular momentum is zero, $L=0$. We implement this by setting the
initial mean occupations: $\langle n_{xy}\rangle =1$, $\langle n_{yz}\rangle
=\langle n_{zx}\rangle =1/2$ (ferro OO), and the off--diagonal elements to
be zero. For this setup, the initial order \textit{does not} persist until
convergence is reached. Instead, there are significant non--zero
off--diagonal elements, on the same order as the occupied diagonal elements,
in the final density matrix. Upon diagonalizing this final matrix, the
orbital order we get has the second electron occupying alternately $\psi
_{+}=(\psi _{yz}+\psi _{zx})/\sqrt{2}$ and $\psi _{-}=(\psi _{yz}-\psi
_{zx})/\sqrt{2}$ along the $c$--axis, which again has the same $I4_{1}/a$
symmetry considered in the preceding subsection. Thus,
we start with an orbital order with $I4_{1}/amd$ symmetry, but the
self--consistent solution breaks certain discrete symmetries and results in
an order with $I4_{1}/a$ symmetry. We thus label this order `AFOO-II'. We note that this order is similar to the one obtained for ZnV$_2$O$_4$ using the same LSDA+$U$ scheme~\cite{maitra07}.

The \textit{collinear} spin fat bands of V $t_{2g}$ and $e_{g}$ are shown in
Fig.~\ref{fig3}(a), and the same for \textit{non-collinear} spins in Fig.~\ref%
{fig3}(b). Qualitative features of the band structures do not change
significantly between the collinear and non--collinear spin configurations. In both plots,
the occupations and dispersions of the $yz$ and $zx$ bands are nearly identical since these orbitals contribute equal weights to the true orbitals, although their relative signs in the linear combinations might differ in these depending on the particular V atom. We also find an insulating band--gap, which in this case is smaller than for `AFOO-I.'

\subsection{Complex Ferro Orbital Order: $I4_{1}/amd$ symmetry}

\begin{figure}[tbp]
\includegraphics[width=1.0\columnwidth]{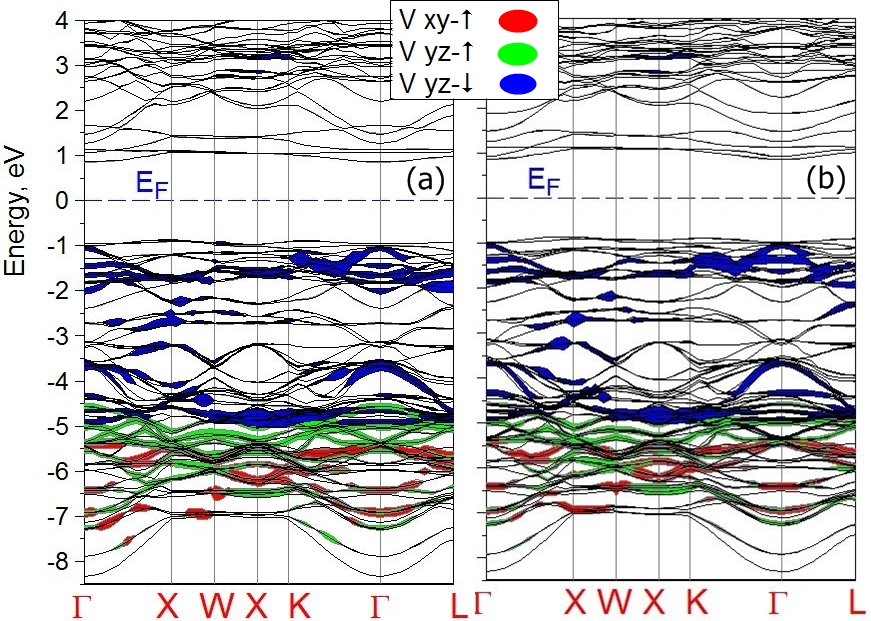}
\caption{(a) The V $t_{2g}$-$\uparrow$ band characters for the low-$T$ tetragonal phase with a \textit{collinear} FEM spin configuration
and complex ferro--orbital order with spin--orbit coupling (SOC-FOO)\ as discussed in text.
(b) Bands for the same setup as in (a) but with a \textit{non-collinear} ferrimagnetic spin configuration.
In both panels, the partial characters of the $xy$-$\uparrow$, $yz$-$\uparrow$, $yz$-$\downarrow$ orbitals are for the sublattice {\rm i, j, k, l} V atoms. There is a band-gap of $E_{\rm gap}=1.76$ eV. The V $xy$-$\downarrow$ band lies above $E_F$, the $zx$-$\uparrow$ bands coincide with the $yz$-$\uparrow$ so we omit it, and finally the $zx$-$\downarrow$ bands are in the same energy region as the $yz$-$\downarrow$ so again we omit it. }
\label{fig4}
\end{figure}

We focus first on the case of small trigonal distortion. The last OO has one electron in $xy$ as before, and the second electron in
the spherical harmonic $L_{z}=-1$, $S_{z}=+1/2$ state, which is a complex
linear combination of $yz$ and $zx$, on all V sites. This is an initial
ferro--orbital order, but with SOC switched on and non--zero orbital angular
momentum. The initial density matrix configuration persists until
convergence. This calculation is carried out using LSDA+$U$+SO. This scenario
is also illustrated by Fig.~\ref{fig1}(b), except that each V atom now carries
a non--zero orbital angular momentum of magnitude one due to the complex
linear combination; hence, there is a uniform orbital order on all V atoms
with $L=1$ in the $3d$ density matrix. The reason for choosing the opposite $%
z$-projections for $\vec{L}$ and $\vec{S}$ is that spin--orbit interaction lowers the
energy for such a setup, compared to the case of having the same sign for
both $z$--projections.

In Fig.~\ref{fig4}(a) we present the band structure of MnV$_{2}$O$_{4}$, for
the \textit{collinear} magnetic configuration, with the V $t_{2g}$-$\uparrow
$ partial characters of the SOC uniform orbital order. We find that the V $%
t_{2g}$-$\downarrow $ and $e_{g}$ characters are above the $E_{F}$ as
expected. For Mn, all the $3d$-$\downarrow $ are below $E_{F}$, while the $%
3d $-$\uparrow $ are above. There is a band gap of $1.76$ eV. Within LSDA+$U$
a half--metallic solution was found in Ref.~\onlinecite{sarkar09}, with only
the $\uparrow $--spin bands of V atoms crossing the $E_{F}$ level, a result
which we have also confirmed \cite{thesis}. Our result is that inclusion of SOC in LSDA+$U$
opens a band gap, signaling a half-metal-to-insulator transition as the SO
coupling parameter is switched on. Since we argue that the uniform complex
ferro order is the correct orbital order based on exchange constant
calculations, we predict a half-metal-to-insulator transition to occur in
single crystalline MnV$_{2}$O$_{4}$ as the temperature goes below $T_{S}$.
In Fig.~\ref{fig4}(b) we present the corresponding band structure of the
\textit{non-collinear} magnetic configuration for this order, with the
partial characters of V $t_{2g}$ shown. We find that the Mn atoms carry no
orbital moment as expected, but the V atoms have an orbital moment $m_{o}=1.03$. The
spin moments are $m_{s}=4.33$ for Mn, and $m_{s}=1.71$ for V. Since the spin
and orbital moments are antiparallel due to SOC coupling, the total moment
for V is $m\approx 0.7$ in this phase.

When we perform the corresponding LSDA+$U$+SO calculation with an $I4_1/a$ symmetry large trigonal distortion, and $U=4.5$ eV, $J_H=1.0$ eV, we find that the converged density matrices are not significantly different from the ones obtained with the small $I4_1/amd$ trigonal distortion, therefore, with respect to the density matrices, the larger trigonal distortion has a minor effect. However, the trigonal distortion does seem to have a rather large effect on the exchange interactions as described further below. The magnetic moments with the larger trigonal distortion are, for Mn atoms: $m_{s}=4.26$, $m_{o}=0.0$ (since the orbital moment is quenched); and for V atoms: $m_{s}=1.65$, $m_{o}=0.87$, giving a total $m=0.78$, similar to what we obtained with a small trigonal distortion. We label the order obtained with spin-orbit coupling as `SOC-FOO'.

\section{Results for Exchange Interactions}

Here we outline the spin--wave model, the ground state spin configuration,
and present the results for our calculated exchange constants $J$ and
single--site anisotropy parameters $D$. Our obtained spin wave spectra of MnV%
$_{2}$O$_{4}$ and comparisons with the neutron scattering experiments are
also given.

\subsection{Spin Wave Model}

The parameters of the model are: [1] the exchange constants $J$ derived from
the LSDA+$U$(+SO) converged charge densities using linear response theory and the
magnetic force theorem \cite{liech87,wan06}, and [2] the single--ion
anisotropy parameters $D$ calculated using an exact--diagonalization atomic
multiplet procedure \cite{alders01}. We input these parameters into the
Heisenberg model Hamiltonian with anisotropy terms, minimize the classical
energy to find the stable ground state configuration, and calculate the
spin--wave excitation spectra. The model Hamiltonian is:

\begin{widetext}
\begin{eqnarray}
H_{\rm spin} &=& - \sum_{\langle {\rm ij} \rangle} J_{\rm ij} \vec{S_{\rm i}} \cdot \vec{S_{\rm j}} - \sum_{\langle {\rm ik} \rangle}J_{\rm ik} \vec{S_{\rm i}} \cdot \vec{S_{\rm k}}
- \sum_{\langle {\rm il} \rangle}J_{\rm il} \vec{S_{\rm i}} \cdot \vec{S_{\rm l}} - \sum_{\langle {\rm jk} \rangle}J_{\rm jk}\vec{S_{\rm j}} \cdot \vec{S_{\rm k}}
- \sum_{\langle {\rm jl} \rangle}J_{\rm jl} \vec{S_{\rm j}} \cdot \vec{S_{\rm l}} - \sum_{\langle {\rm kl} \rangle}J_{\rm kl}\vec{S_{\rm k}} \cdot \vec{S_{\rm l}} \nonumber
\\
&-& J_{\rm Mn-V}\sum_{\langle {\rm (p,q)(i,j,k,l)} \rangle}(\vec{S_{\rm p}} + \vec{S_{\rm q}}) \cdot (\vec{S_{\rm i}}+\vec{S_{\rm j}}+\vec{S_{\rm k}}+\vec{S_{\rm l}}) - \sum_{\langle {\rm pq} \rangle}J_{\rm pq} \vec{S_{\rm p}} \cdot \vec{S_{\rm q}}
+ \sum_{\rm x=i,j,k,l,p,q}\vec{S_{\rm x}} \cdot \bar{D}_{\rm x} \cdot \vec{S_{\rm x}}.
\label{eqn1}
\end{eqnarray}
\end{widetext}

The subscripts on the $J$ label the four inequivalent V sublattices, $\mathrm{i}
$, $\mathrm{j}$, $\mathrm{k}$, $\mathrm{l}$, and two inequivalent Mn
sublattices, $\mathrm{p}$, $\mathrm{q}$. The $J_{\mathrm{Mn-V}}$ is taken
outside the summation because it has the same value for all pairs of Mn and
V atoms. All the $J$ couplings are between nearest--neighbor atoms of two
different sublattices, and each pair is counted only once in the summation
over all sites. We ignore the next--nearest--neighbor couplings because we
found them to be much smaller in magnitude.

\subsection{Spin Configuration}

\begin{figure}[h!]
\includegraphics[width=0.9\columnwidth]{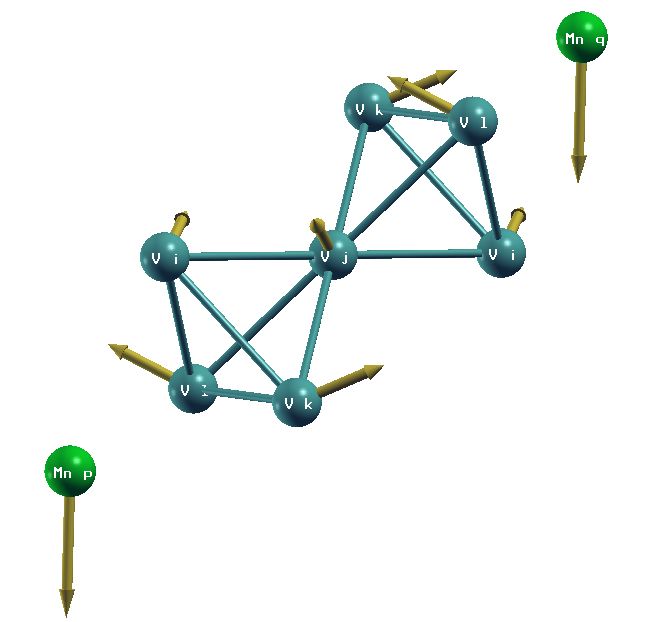}
\caption{The $T=0$ non--collinear spin configuration of the V spins from
Ref.~\onlinecite{gar08}.}
\label{fig5}
\end{figure}

The low--$T$ spin structure is non--collinear for V atoms, and collinear for
Mn atoms with respect to the $c$--axis. The pyrochlore lattice on which the
V atoms sit is geometrically frustrated for nearest--neighbor isotropic ($%
J^{ab}=J^{c}$) AFM exchange. The frustrated pyrochlore interactions mean
that there could be a macroscopic ground state degeneracy. But this
frustration is partially relieved in the low--$T$ phase by the presence of
additional nearest--neighbor exchange interactions with Mn atoms, tetragonal
distortion, and orbital ordering. The last one has the effect of making the
V--V AFM exchange anisotropic: $J^{ab}\neq J^{c}$. It is well--known that
the orbital or magnetic degeneracy can be lifted by the coupling of these
degrees of freedom with the lattice via the Jahn--Teller effect \cite%
{jahn37,tokura00}. The ground--state spin configuration selected by the
system in the low--$T$ phase is non--collinear due to the combined effect of
the frustration and coupling of the V spins to Mn spins, V $t_{2g}$
orbitals, and the lattice. In this structure, the V--atom spins develop
components in the $ab$ plane perpendicular to each other. The amount of
canting away from the $c$--axis can be characterized by a single canting
angle $\theta $. Given the values of all $J$ and $D$ in Eq.~\ref{eqn1}, one
can find the angle $\theta $ as a function of $J$ and $D$ that will minimize
the classical ground--state energy of the configuration (derivation given in
Ref.~\onlinecite{thesis}):

\begin{equation}
\theta =\arccos [-\frac{3J_{\rm Mn-V}S_{\rm Mn}}{%
(D_{\rm V}^{z}-D_{\rm V}^{x,y}-2J_{\rm V-V}^{c}-2J_{\rm V-V}^{ab})S_{\rm V}}].  \label{eqn2}
\end{equation}
The non--collinear spin configuration that achieves this energy minimum is
shown in Figure~\ref{fig5} \cite{xcrysden}.

\subsection{Exchange Constants}

\begin{table*}[!]
\begin{tabular}{|c|c|c|c|c|c|c|}
\hline
meV & No OO $U=0$ eV & No OO $U=5$ eV & AFOO-I $U=5$ eV & AFOO-II $U=5$ eV &
SOC-FOO $U=5$ eV & Expt.\cite{chung08} \\ \hline
$J_{\mathrm{ii}}$ & -2.72 & 0.136 & 0.3264 & -0.04488 & 0.1496 & -
\\ \hline
$J_{\mathrm{V-V}}^{ab}$ & -20.4 & -21.76 & -14.96 & -19.04 & -10.88 &
-9.89 \\ \hline
$J_{\mathrm{V-V}}^{c}$ & -20.4 & -18.36 & -3.536 & -7.072 & -2.72 & -3.08
\\ \hline
$J_{\mathrm{Mn-V}}$ & -10.2 & -2.992 & -5.44 & -5.44 & -4.76 & -2.82
\\ \hline
$J_{\mathrm{pq}}$ & 1.2 & 2.167 & 2.72 & 2.72 & 2.72 & - \\ \hline
$J_{\mathrm{pp}}$ & -0.476 & 0.204 & 0.204 & 0.272 & 0.272 & - \\
\hline
\end{tabular}%
\caption{Calculated exchange constants in meV for the \textit{collinear}
ferrimagnetic configurations and imposing various orbital orders along with the $I4_1/amd$ small trigonal distortion. The last
two columns list our theoretical $J$s for the spin--orbit coupled
ferro--orbital order (SOC--FOO) and experimental $J^{\prime }$s from Ref.~
\onlinecite{chung08} respectively. The experimental $J^{\prime }$s match the
SOC--FOO $J^{\prime }$s better than for the other theoretical $J^{\prime }$%
s. }
\label{table1}
\end{table*}

\begin{table*}[!]
\begin{tabular}{|c|c|c|c|c|c|}
\hline
meV & $U=4.5$ eV & $U=5.0$ eV & $U=5.5$ eV & $U=6.0$ eV & Expt.\cite{chung08} \\ \hline
$J_{\mathrm{ii}}$ & 0.449 & 0.35 & 0.3 & 0.272 & -
\\ \hline
$J_{\mathrm{V-V}}^{ab}$ & -17.7 & -14.28 & -12.92 & -11.56 & -9.89
 \\ \hline
$J_{\mathrm{V-V}}^{c}$ & -4.624 & -4.352 & -3.808 & -3.4 & -3.08
\\ \hline
$J_{\mathrm{Mn-V}}$ & -6.8 & -5.712 & -5.304 & -4.896 & -2.82
\\ \hline
$J_{\mathrm{pq}}$ & 2.72 & 2.584 & 2.448 & 2.312 & - \\ \hline
$J_{\mathrm{pp}}$ & 0.204 & 0.2 & 0.2 & 0.1768 & - \\
\hline
\end{tabular}%
\caption{Calculated exchange constants in meV for the \textit{collinear}
ferrimagnetic configurations with the $I4_1/a$ large trigonal distortion of the VO$_6$ octahedra for U=4.5, 5.0, 5.5, 6.0 eV. In all these cases, the self--consistency converges to the orbital order named `AFOO-I', which is also the same orbital order found in Ref.~\onlinecite{sarkar09}. For comparison the last column lists the experimental values. $U=6$ eV gives theoretical $J^{\prime}$s which approach the experimental $J^{\prime}$s.}
\label{table3}
\end{table*}

\begin{table*}[!]
\begin{tabular}{|c|c|c|c|c|c|}
\hline
meV & $U=4.5$ eV & $U=5.0$ eV & $U=5.5$ eV & $U=6.0$ eV & Expt.\cite{chung08} \\ \hline
$J_{\mathrm{ii}}$ & 0.272 & 0.204 & 0.177 & 0.15 & -
\\ \hline
$J_{\mathrm{V-V}}^{ab}$ & -15.64 & -12.24 & -10.61 & -9.11 & -9.89
 \\ \hline
$J_{\mathrm{V-V}}^{c}$ & -5.8 & -4.352 & -3.536 & -2.788 & -3.08
\\ \hline
$J_{\mathrm{Mn-V}}$ & -6.12 & -5.44 & -4.896 & -4.352 & -2.82
\\ \hline
$J_{\mathrm{pq}}$ & 2.72 & 2.72 & 2.45 & 2.329 & - \\ \hline
$J_{\mathrm{pp}}$ & 0.272 & 0.272 & 0.231 & 0.231 & - \\
\hline
\end{tabular}%
\caption{Calculated exchange constants in meV for the \textit{collinear}
ferrimagnetic configurations with the $I4_1/a$ large trigonal distortion of the VO$_6$ octahedra for U=4.5, 5.0, 5.5, 6.0 eV \textit{and} SOC. In all these cases, the self--consistency converges to the orbital order named `SOC-FOO'. For comparison the last column lists the experimental values. Again, the theoretical $J^{\prime}$s for $U=6$ eV come closest to experiment.}
\label{table4}
\end{table*}

In Table~\ref{table1} we present the $J$ parameters that we calculate using
LSDA+$U$(+SO)\ method and magnetic force theorem for the small trigonal distortions and the indicated $U$ values. In Tables~\ref{table3} and~\ref{table4} we present the $J^{\prime}$s for the large trigonal distortion: Table~\ref{table3} for `AFOO-I' and Table~\ref{table4} for `SOC-FOO'. As the method computes the
exchange constants in reciprocal space, we fourier transform them and show
only nearest--neighbor exchange interactions between atoms of each
sublattice. For the spinel structure, any of nearest--neighbor pairs always
belongs to a different sublattice. The values of $J_{\rm V-V}$ for no orbital
order and $U=0$ eV are the same for all V-V pairs; but when $U=5$ eV,
there is a tendency for anisotropy to develop: the in--plane $J_{\rm V-V}^{ab}$
becomes unequal from the out--of--plane $J_{\rm V-V}^{c}$. This shows that the anisotropy in the $J_{\rm V-V}$, and
the orbital ordering which causes it, could both be interaction driven. When
there is an orbital order, $J_{\rm V-V}$ is different along the $ab$ V chains
and between the chains (along the $c$--axis), as expected. This is true even
for the case of the uniform orbital orders, because the exchange matrix
elements of the Coulomb operator will be different within the $ab$ plane and
between the planes, as can be seen from the shapes of the occupied orbitals
in Fig.~\ref{fig1}(a,b).

We have calculated the exchange constants within LSDA+$U$ and LSDA+$U$+SO taking into account both small and large trigonal distortions with $I4_1/amd$ and $I4_1/a$ symmetry respectively. The larger trigonal distortion result in $J^{\prime}$s that are ~$50\%-80\%$ larger, with and without SOC. Thus the effect of increasing trigonal distortion keeping $U=4.5$ eV affects the agreement with the experimental spin-waves. One explanation for why this is so could be that $U=4.5$ eV is too small to describe the correlation effects in V. In order to check this, we also tried $U=4.5$, $5.0$, $5.5$, $6.0$ eV for large trigonal distortions and found that the $J^{\prime}$s indeed decrease as $U$ increases, see Tables~\ref{table3},~\ref{table4}. By varying both the size of the trigonal distortion and the $U$, we are dealing with a two-parameter problem. Since neither the trigonal distortion nor $U$ are exactly known, we have presented our results as an exploration of the trends in $J^{\prime}$s within this two-parameter space. The increase in $U$ will bring down the values of the $J^{\prime}$s as they typically scale as $t_{dd\sigma}^2/U$ for direct exchange between the V atoms. The trends in the variation of $J^{\prime}$s within the two-parameter space indicate that the $J^{\prime}$s for SOC-FOO best describes the experimental spin waves for small trigonal distortion with $U=5$ eV, Table~\ref{table1}, and larger trigonal distortion with $U=6$ eV, Table~\ref{table4}.

\subsection{Single--Ion Anisotropy}

\begin{table}[!]
\begin{tabular}{|c|c|c|c|c|}
\hline
- & CF $E_{xy}$ eV & CF $E_{e_g}$ eV & Theory meV & Expt.\cite{chung08} meV
\\ \hline
Mn $D^z$ & -0.016 & 1.0 & -0.1123 & -0.1024 \\ \hline
V $D^{x,y}$ & -0.024 & 0.4 & -4.056 & -4.04 \\ \hline
V $D^z$ & -0.024 & 0.4 & 7.34 & 2.79 \\ \hline
\end{tabular}%
\caption{Table of calculated anisotropy constants for V$^{3+}$ $3d^{2}$ and
Mn$^{2+}$ $3d^{5}$ atomic shells. The energies of $E_{xy}$ and $E_{e_g}$ due to the crystal--fields
are measured with respect to $E_{yz,zx}=0$ eV. }
\label{table2}
\end{table}

The calculation of single--ion anisotropy requires first the total energies
of the interacting atomic shell in a crystal--field environment along with
the spin--orbit coupling. The method for its computation is described in
Ref.~\onlinecite{alders01, thesis}, so here we merely present our results.
The input parameters used in the total energy calculation are: the SOC parameter
$0.15$ eV, and Slater integrals $F_{0}=5.0$ eV, $F_{2}=7.6$ eV, $F_{4}=4.7$ eV.
We then vary the direction of the magnetic moment by applying a small
external magnetic field. The CF levels have the following values: the energies of $yz$ and $%
zx$ are both set to the reference value of $0.0$ eV. The $e_{g}$ level is
varied from $0.2$ eV to $1.0$ eV in steps of $0.2$ eV; and the $xy$ level
has the energies $-0.024$ eV, $-0.016$ eV, $-0.008$ eV, $0.0$ eV. This gives
a set of 20 different CF configurations. The $E_{xy}=E_{yz/xz}$ represents
cubic CF, and $E_{xy}\neq E_{yz/xz}$ represents tetragonal CF.

The total atomic shell energies thus obtained are fitted to a parabolic
function of the polar angle $\theta $ representing moment orientation,
centered at $\theta =0$ in the case of $z$--axis anisotropy, and centered at
$\theta =\pi /2$ in the case of $x/y$--axis anisotropy. The results of the
parabolic fit that best match the experimentally known $D$ values are given
in Table~\ref{table2} for both the V $3d^{2}$ and Mn $3d^{5}$ shells.

The easy axis for Mn is $z(c)$--axis, and for V it is either $x$ or $y$. The
easy axis always has a negative anisotropy parameter, which means the energy
is lowered when the spin projection along the easy axis is maximized. For
Mn, the spin projection along $z$ tends to be maximized. However, V also has
a positive anisotropy parameter along the $z$--axis. So, V spin projection
likes to be maximized along $y$, and minimized along $z$\cite{chung08}.
Thus, the V spin moment has a tendency to be in a non--collinear direction
with respect to the $z$--axis. Our anisotropy computation is able to
reproduce these signs as well as magnitudes for Mn ($3d^{5}$) and V ($3d^{2}$%
) shells.

Looking at the anisotropy fit values we see that the value of Mn anisotropy
reported in Ref.~\onlinecite{chung08} are obtained for $E_{xy}=-0.016$ eV, $%
E_{e_{g}}=1.0$ eV, namely $D_{\rm Mn}^{z}=-0.1123$ meV, which is similar to the
literature value. For our fitted values of V anisotropy, we do not find such
a close match, but there are several CF values which give the anisotropy of
Ref.~\onlinecite{chung08} up to the correct sign and order of magnitude. For
example, $E_{xy}=-0.024$ eV, $E_{e_{g}}=0.4$ eV give $D_{\rm V}^{z}=7.34$ meV,
and $D_{\rm V}^{x,y}=-4.056$ meV, which can be compared to $D_{\rm V}^{z}=2.79$ meV,
and $D_{\rm V}^{x,y}=-4.04$ meV of Ref.~\onlinecite{chung08}. One reason why our
calculated $D_{\rm V}^{z}$ parameter differs by a large amount from the
experimental value is because we have to tune the CF energy levels to
simultaneously match two different single--ion anisotropies, and it was not
possible to get them both to match the experimental $D$ values of V.

\subsection{Spin--Wave Spectra}

\begin{figure}[!]
\includegraphics[width=0.83\columnwidth]{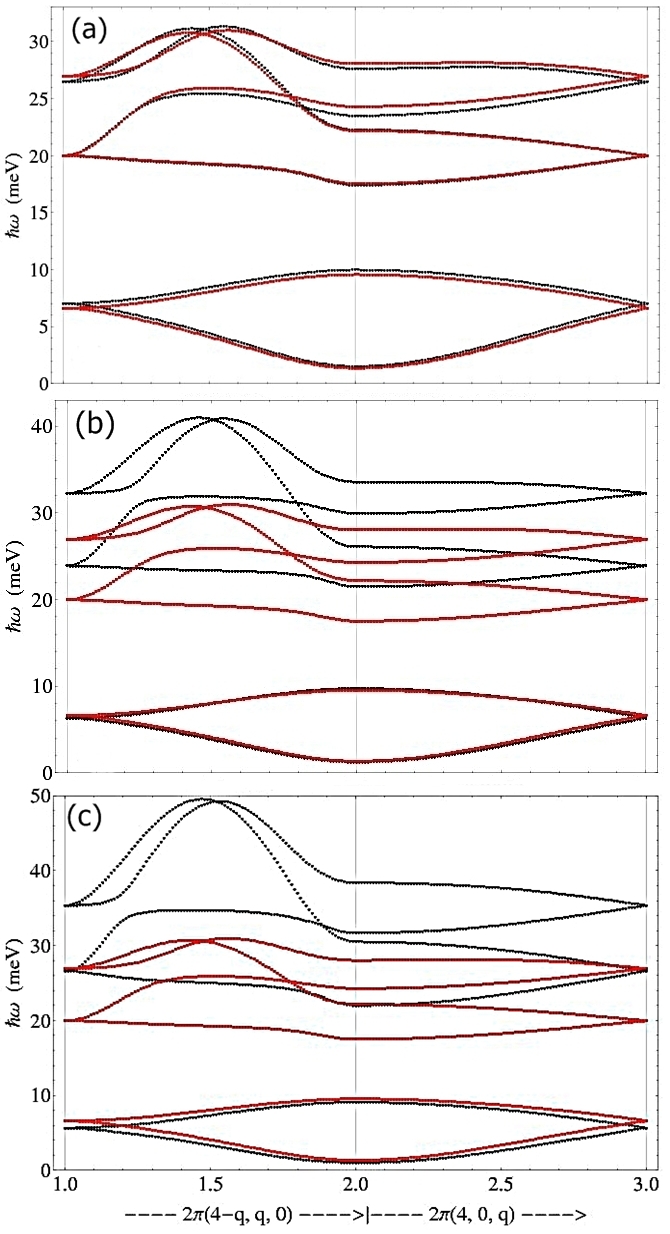}
\caption{In all panels, the red lines are experimental and black
lines are theoretical spin--waves: (a) Spin wave spectrum for $I4_{1}/amd$ spin--orbit coupled
ferro--orbital order (SOC--FOO)\ along the high--symmetry lines of the
Brillouin zone. We find an excellent match between our
theoretical and previous experimental data of Ref.~\onlinecite{chung08}.
(b) Spin--waves corresponding to the $I4_{1}/a$ symmetry AFOO--I order.
The upper four V oscillation branches of the theoretical spin--waves are both
too high in energy and have a larger dispersion compared to the experimental plot.
(c) Same as in (b), but for the AFOO--II order. Here the overestimate in the $J_{\rm V-V}$ is even greater than in (b). All theoretical spin-wave plots are for $U=5$ eV and small trigonal distortion.}
\label{fig6}
\end{figure}

We developed a code to compute the linear spin wave spectra for the
non--collinear spin configuration. The program takes as an input our
computed values of $J$ and $D$. We first find the ground state which will in
general be a non--collinear configuration with the spins pointing along the
local quantization axis as given by $\theta $ in Eq.~\ref{eqn2}. We second
find the Heisenberg equations of motion, and numerically diagonalize the
resulting system of linear equations. The resulting spin--wave spectra are
plotted in Fig.~\ref{fig6}(a) for the `SOC-FOO' uniform orbital order, along with the
experimental spin waves. We find that the spin waves obtained from the $J$
and $D$ values of the SOC ferro--orbital order with small trigonal distortion and $U=5$ eV matches well with experiment, although other combinations of trigonal distortion and $U$ could also yield similar $J^{\prime}$s.
We also note that the lower two modes are due to the oscillations of Mn
spins: The lower energy being the symmetric mode, and the higher energy the
anti--symmetric mode\cite{chung08}. The upper four modes are oscillations of
the V spins\cite{chung08}.

For comparison, we show the spin waves for the other orbital orders, also obtained with small trigonal distortion and $U=5$ eV,
that \textit{do not} match well with the experimental data. The model parameters
for these orbital orders do give a reasonable spin canting angle when using
Eq.~\ref{eqn2}, but the upper branches of the spin waves corresponding to
the V oscillations are too high in energy and have a larger band--width in these plots (due to
considerable overestimate of V--V exchange) compared to the correct one in
Fig.~\ref{fig6}(a). Figure~\ref{fig6}(b) shows the spin waves for the `AFOO-I' order.

Figure~\ref{fig6}(c) shows the spin waves for the `AFOO-II', which is composed
of real linear combinations of $yz$ and $zx$ orbitals with the relative sign
between $yz$ and $zx$ alternating between $ab$ layers along the $c$ axis. As
Table~\ref{table1} shows, this order again gives a considerably greater V--V
exchange compared to the experiment and therefore the upper branches are
much higher in energy and have a greater dispersion relative to the experimental plot.

We conclude that the excellent agreement between our theoretical and
experimental spin--wave dispersions for all the six oscillation modes can be
obtained for a sample setup with SOC ferro--orbital order with $I4_{1}/amd$ small trigonal distortion, where the second $t_{2g}$
electron occupies a complex linear combination of $|yz\rangle \pm
i|zx\rangle $ uniformly on all V--sites. The incorporation of the low symmetry $I4_1/a$ large trigonal distortion tends to increase the $J_{\rm V-V}$'s by 50$\%$-80$\%$, but we find that by a reasonable increase of the Coulomb parameter to $U=6$ eV, we can obtain $J^{\prime}$s that match the experimental ones. The trend we notice is that small trigonal distortion + lower $U$ as well as large trigonal distortion + higher $U$ both give $J^{\prime}$s that are close to the experimental $J^{\prime}$s, however, the former case with `SOC-FOO' seems to give the best match of all the combinations we have tried. The other two orbital orders, `AFOO-I' and `AFOO-II', do not give such a good match with experiment throughout the Brillouin zone for the same value of distortion and $U$ so these orders may be ruled out. We further note that the spin--orbit
coupling plays an important role in the orbital physics of V--atoms in MnV$%
_{2}$O$_{4}$. This is also justified by the fact that the single--ion
anisotropy is relatively high, as evidenced by the large gaps for the
would--be acoustic modes at $\Gamma$.

\section{Conclusion}

By theoretical computations of the interatomic exchange constants using
LSDA+$U$(+SO) method, magnetic force theorem and by imposing various orbital
ordering scenarios we have shown that the orbital order on the V sites of
MnV$_{2}$O$_{4}$ is similar to a complex linear combination of $zx$ and $yz$
on all V sites. Our calculated spin wave spectra for this order come closest
to the experimental data. Further support in evidence of the complex
order is the strong single--ion anisotropy experienced by the spin moments
on the V sites, as well as the \textit{reduction} of the V magnetic moment in the low--$T$ phase\cite{gar08} which could not be captured by LSDA+$U$ alone. We also predict, based on our $U=5$ eV, orbital--ordered band--structures,
that the low--$T$ phase of MnV$_2$O$_4$ is a Mott--type insulator, and that a
half--metal--to--insulator transition accompanies the simultaneous orbital ordering,
structural distortion, and non-collinear moment transitions at $T_S=53$ K.

\textbf{Acknowledgements.}

The authors acknowledge useful discussions with Myung Joon Han, Rajiv Singh and Nick Curro.
The work was supported by DOE SciDAC Grant No. SE-FC02-06ER25793 and by DOE
Computational Material Science Network (CMSN) Grant No. DE-SC0005468.


\begin{thebibliography}{99}
\bibitem{tokura00} Y. Tokura and N. Nagaosa, Science \textbf{288}, 462,
(2000).

\bibitem{plum87} R. Plumier and M. Sougi, Solid State Commun. \textbf{64},
53 (1987); Physica B \textbf{155}, 315 (1989).

\bibitem{gar08} V. O. Garlea, R. Jin, D. Mandrus, B. Roessli, Q. Huang, M. Miller, A. J. Schultz, and S. E. Nagler, Phys. Rev. Lett. \textbf{100},
066404 (2008).

\bibitem{sarkar09} S. Sarkar, T. Maitra, Roser Valent\'{\i}, and T.
Saha-Dasgupta, Phys. Rev. Lett. \textbf{102}, 216405 (2009).

\bibitem{tsun03} H. Tsunetsugu and Y. Motome, Phys. Rev. B \textbf{68},
060405(R) (2003).

\bibitem{suzuki07} T. Suzuki, M. Katsumura, K. Taniguchi, T. Arima, and T.
Katsufuji, Phys. Rev. Lett. \textbf{98}, 127203 (2007).

\bibitem{adachi05} K. Adachi, T. Suzuki, K. Kato, K. Osaka, M. Takata, and T.
Katsufuji, Phys. Rev. Lett. \textbf{95}, 197202 (2005).

\bibitem{tchern04} O. Tchernyshyov, Phys. Rev. Lett. \textbf{93}, 157206
(2004).


\bibitem{jahn37} H. A. Jahn and E. Teller, Proc. R. Soc. A \textbf{161}, 200
(1937).

\bibitem{DFTBook} For a review, see, e.g., Theory of the Inhomogeneous
Electron Gas, edited by S. Lundqvist and S. H. March (Plenum, New York,
1983).

\bibitem{AnisimovLDA+U} {For a review, see, e.g., \emph{Strong Correlations
in electronic structure calculations}}, edited by V.~I. Anisimov (Gordon and
Breach Science Publishers, Amsterdam, 2000).

\bibitem{AndersenNMTO} O. K. Andersen, T. Saha--Dasgupta, Phys. Rev. B
\textbf{62}, 16219, (2000).

\bibitem{chung08} J.-H. Chung, J.-H. Kim, S.-H. Lee, T. J. Sato, T. Suzuki, M. Katsumura, and T. Katsufuji, Phys. Rev. B \textbf{77},
054412 (2008).

\bibitem{alders01} D. Alders, R. Coehoorn, W. J. M. de Jonge, Phys. Rev. B
\textbf{63}, 054407 (2001).

\bibitem{good63} J. B. Goodenough, \textit{Magnetism and the Chemical Bond}
(Interscience, New York, 1963); J. Kanamori, J. Phys. Chem. Solids \textbf{10%
}, 87 (1959).

\bibitem{baek09} S.-H. Baek, N. J. Curro, K.-Y. Choi, A. P. Reyes, P. L. Kuhns, H. D. Zhou, and C. R. Wiebe, Phys. Rev. B \textbf{80},
140406(R) (2009).

\bibitem{Perkins} Gia-Wei Chern, Natalia Perkins, and Zhihao Hao, Phys. Rev.
B 81, 125127 (2010).

\bibitem{Andersen1975} O. K. Andersen, Phys. Rev. B \textbf{12}, 3060 (1975).

\bibitem{Savrasov1996} S. Y. Savrasov, Phys. Rev. B \textbf{54}, 16470
(1996).

\bibitem{liech87} A. I. Liechtenstein, M. I. Katsnelson, V. P. Antropov, and
V. A. Gubanov, J. Magn. Magn. Mater. 67, 65 (1987).

\bibitem{wan06} X. Wan, Q. Yin, and S. Y. Savrasov, Phys. Rev. Lett. \textbf{97},
266403 (2006).



\bibitem{thesis} R. Nanguneri, Ph.D. thesis, (2012).

\bibitem{miyake08} T. Miyake and F. Aryasetiawan, Phys. Rev. B \textbf{77},
085122 (2008).

\bibitem{maitra07} T. Maitra and R. Valent\'{\i}, Phys. Rev. Lett. \textbf{99}, 126401 (2007).

\bibitem{xcrysden} A. Kokalj, Comp. Mater. Sci., \textbf{2003}, Vol. 28, p. 155.
\end{thebibliography}
\end{document}